\documentclass[aps,prl,twocolumn,superscriptaddress,showpacs]{revtex4}
\usepackage{graphics}
\usepackage{bm}

\begin{document}

\title{Persistent current in superconducting nanorings}

\author{K.A. Matveev} 

\affiliation{Department of Physics, Duke University, Box 90305, Durham, NC
  27708}

\author{A.I. Larkin} 
\author{L.I. Glazman}

\affiliation{Theoretical Physics Institute, University of Minnesota, 
Minneapolis, MN 55455}

\date{July 10, 2002}

\begin{abstract}
  The superconductivity in very thin rings is suppressed by quantum phase
  slips.  As a result the amplitude of the persistent current oscillations
  with flux becomes exponentially small, and their shape changes from
  sawtooth to a sinusoidal one.  We reduce the problem of low-energy
  properties of a superconducting nanoring to that of a quantum particle
  in a sinusoidal potential and show that the dependence of the current
  on the flux belongs to a one-parameter family of functions obtained by
  solving the respective Schr\"odinger equation with twisted boundary
  conditions.
\end{abstract}


\maketitle

The properties of superconducting grains change dramatically when their
size shrinks to a few nanometers.  This phenomenon has recently generated
a lot of interest~\cite{ralph}.  In particular, it has been established
both theoretically and experimentally that the superconductivity cannot be
observed if the quantum level spacing in a nanoparticle exceeds the
superconducting gap $\Delta$.  The recent success in manufacturing of
superconducting nanowires \cite{bezryadin1} has raised similar questions
regarding the superconductivity in one-dimensional objects.  It has been
demonstrated experimentally \cite{bezryadin2} that as the nanowire becomes
thiner, the superconducting transition in it disappears, and a finite
resistance is observed at low temperatures.  The suppression of the
 superconductivity in thin wires was attributed \cite{bezryadin2} to the
destruction of the phase coherence by quantum phase slips
\cite{zaikin,golubev}. 

To understand the superconducting properties of thin wires one has to keep
in mind that in an infinite one-dimensional conductor all the electronic
states are localized.  The localization length $\lambda$ in a wire of
cross-section $A$ can be estimated as $\lambda\sim k_F^2 Al$, where $k_F$
is the Fermi wavevector and $l$ is the mean free path.  If the length $L$
of the wire is shorter than $\lambda$, it can be viewed as a metal grain
which becomes a good superconductor if the gap $\Delta$ exceeds the
quantum level spacing $E_F/k_F^3AL$.  On the other hand, a wire of length
$L\gg\lambda$ cannot be viewed as a good conductor, and its
superconducting properties may be affected by localization.  

If the attractive interactions between the electrons are weak, so that the
bulk superconducting gap $\Delta$ is small compared to the level spacing
$\delta_\lambda\sim E_F/k_F^3A\lambda$ of a piece of the wire of length
$\lambda$, the superconductivity is suppressed on a microscopic scale.  In
the more interesting regime of $\Delta\gg \delta_\lambda$, the
superconducting gap is not affected by the localization \footnote{The
  ratio $\delta_\lambda/\Delta$ coincides with the Ginzburg parameter
  ${\rm Gi}$.  In this paper we concentrate on the regime of ${\rm
    Gi}\ll1$.}.  The important question in this regime is that of phase
coherence between the different parts of the wire.  Experimentally this
issue can be studied by measuring the persistent current in a nanowire
ring as a function of the magnetic flux piercing it.  The magnitude of the
persistent current oscillations as a function of the flux is a direct
measure of the superconducting phase coherence throughout the wire.

\begin{figure}[tb]
 \resizebox{.38\textwidth}{!}{\includegraphics{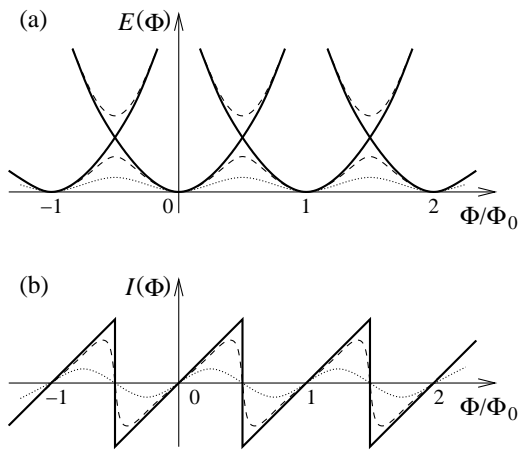}}
\caption{\label{fig:parabolas} 
  (a) The energy of a nanoring as a function of the flux $\Phi$ through
  it.  In the limit of relatively large cross-section $A$ the energy shows
  the classical behavior (\ref{eq:E_classical}) shown by solid line.
  Quantum phase slips result in level splitting shown in dashed line, and
  then lead to sinusoidal dependence of the ground state energy on the
  flux (dotted line). (b) The persistent current in the nanoring,
  $I(\Phi)=c\,dE/d\Phi$.  In the classical limit the current shows
  sawtooth behavior shown by solid line.  As the wire becomes thiner, the
  sawtooth is rounded and eventually transforms to a sinusoidal
  oscillation.}
\end{figure}

In this paper we study the dependence of the persistent current in a
superconducting nanoring as a function of the flux and the size of the
ring.  If the wire is relatively thick, the electrons move freely between
different parts of the wire, and therefore the superconducting phase
$\phi$ is a well defined classical variable.  At low temperatures
$T\ll\Delta$ there are no quasiparticle excitations in the nanoring.  The
allowed states of the system differ by the phase change
$\varphi=\phi(L)-\phi(0)$ accumulated over the circumference of the ring.
At a given value $\Phi$ of the magnetic flux through the ring $\varphi$
can assume the values $\varphi_m=2\pi\Phi/\Phi_0 +2\pi m$, where $m$ is an
arbitrary integer, and $\Phi_0=\pi\hbar c/e$ is the superconducting flux
quantum.  The energies of these states are given by
\begin{equation}
  \label{eq:E_classical}
  E_m=\frac{2\pi^2\hbar^2 n_sA}{m^*L}
      \left(\frac{\Phi}{\Phi_0} + m\right)^2.
\end{equation}
Here $n_s$ is the density of superconducting electrons, $m^*$ is the
electron mass.  The dependences of the energy levels
(\ref{eq:E_classical}) on the flux are shown by solid line in
Fig.~\ref{fig:parabolas}(a).  The persistent current can be found as a
derivative of the ground state energy $I=c\,dE/d\Phi$ and shows the
characteristic sawtooth behavior, Fig.~\ref{fig:parabolas}(b).

The above picture fails in thin wires, where the fluctuations of the
superconducting order parameter cannot be neglected.  The most
important type of the fluctuations is the quantum phase slip
\cite{zaikin,golubev} that changes the phase $\phi$ at a point $x$ by
$\pm 2\pi$.  We will show that the effect of the rare phase slips on
the system reduces to quantum transitions between the levels
(\ref{eq:E_classical}).  Such transitions are most important near the
degeneracy points at half-integer values of $\Phi/\Phi_0$ and result
in the small level splitting shown by dashed line in
Fig.~\ref{fig:parabolas}(a).  We will also show that the multiple
quantum phase slips in longer and thiner wires
eventually lead to a crossover from sawtooth to
sinusoidal behavior of the persistent
current, see dotted line in Fig.~\ref{fig:parabolas}(b).

The theory of quantum phase slips in nanowires \cite{zaikin,golubev}
is rather complex.  In particular, the most important quantitative
parameter describing the phase slip---its action---can be found
analytically only up to an unknown numerical coefficient.  On the
other hand, the effect of the phase slips on the ground state of the
nanoring can be studied using a much simpler model of a chain of
coupled superconducting grains considered below.  Similarly to the
nanowire, the chain of grains will have quantum phase slips that
affect the persistent current in the same way.  We will discuss the
relation between the two models in more detail below.

We consider a model of the chain of superconducting grains defined by the
following imaginary-time action:
\begin{equation}
  \label{eq:action}
  S=\int_0^\beta dt \sum_{n=1}^N\left\{
    \frac{\dot\theta_n^2}{2E_C} + E_J[1-\cos\theta_n(t)]\right\}.
\end{equation}
The grains are assumed to be connected by tunnel junctions with Josephson
coupling energy $E_J$; the capacitance of each junction $C$ gives rise to
the charging energy $E_C=(2e)^2/C$; the variable $\theta_n$ is the phase
difference across the $n$-th junction; $\beta=1/T$.  To model a closed
chain pierced by flux $\Phi=(\varphi/2\pi)\Phi_0$, one should impose an
additional condition
\begin{equation}
  \label{eq:constraint}
   \sum_{n=1}^N \theta_n(t) = \varphi
\end{equation}
on the phases $\theta_n$.  Note that the action (\ref{eq:action}) does not
include the charging energy terms due to the self-capacitance of the
grains.  These terms, characterized by charging energy ${\tilde E}_C$ per
grain, are known \cite{bradley} to give rise to a Kosterlitz-Thouless type
quantum phase transition in an infinite chain of Josephson junctions at
$E_J/{\tilde E}_C\sim1$.  The use of model (\ref{eq:action}) is motivated
by the fact that the electric field is very well screened in metal wires
which corresponds to the condition $\tilde E_C\gg E_J, E_C$.

The ground state properties of the chain can be derived from the partition
function $Z=\int e^{-S} D\theta_n$ in the limit $\beta\to\infty$.  In this
paper we consider the case of strong coupling between the grains $E_J\gg
E_C$.  At $E_C=0$ the phases $\theta_n$ become classical variables, and
the energy states of the chain can be found by minimizing the sum of the
Josephson energy terms in the action (\ref{eq:action}) with the constraint
(\ref{eq:constraint}).  At large number of contacts $N\gg1$ one finds
\begin{equation}
  \label{eq:E_classical_chain}
  E_m=\frac{E_J}{2N} \left(\varphi + 2\pi m\right)^2,
\end{equation}
in complete analogy with Eq.~(\ref{eq:E_classical}) for a metal ring.

At finite small $E_C$ the fluctuations of $\theta_n(t)$ should be taken
into consideration.  First, one can include the Gaussian fluctuations
around the classical solutions $\theta_n(t)=\textrm{const}$.  This is
accomplished by expanding the cosine in the action (\ref{eq:action}) up to
quadratic terms in $\theta_n$.  The resulting correction to the ground
state energy $\delta E_m= (N-1)\sqrt{E_JE_C}$ accounts for the zero-point
oscillations of the particles in the minima of the cosine potential.  This
constant correction does not affect the shape of the oscillations of the
persistent current $I=(2e/\hbar)dE/d\varphi$.  One can also include the
anharmonic terms of the expansion of the cosine potential in the action
(\ref{eq:action}).  The resulting corrections yield a small in $E_C/E_J$
distortion of the sawtooth dependence $I(\Phi)$, but do not significantly
affect the discontinuities at half-integer $\Phi/\Phi_0$.

A more interesting fluctuation of the phases $\theta_n(t)$ involves an
instanton (quantum phase slip), i.e., a trajectory that begins near
one of the minima (\ref{eq:E_classical_chain}) of the 
potential energy in action~(\ref{eq:action}) at $t=0$ and ends near another
minimum at $t=\beta$.  For instance, a trajectory starting at
$\theta_n(0)=\varphi/N$ and ending at
$\theta_n(\beta)=(\varphi-2\pi)/N + 2\pi\delta_{nk}$ for arbitrary $k$
in the interval $1\leq k\leq N$ connects the minima
(\ref{eq:E_classical_chain}) with $m=0$ and $m=-1$.  The shape of the
instanton trajectory can be found by minimizing the classical action
(\ref{eq:action}) with the above boundary conditions on $\theta_n(t)$.
In the case of large number of junctions $N\gg1$ the dominant
contribution is due to the contact $k$ where the phase slip occurs,
and the contributions of the other contacts can be neglected.  Then
one obtains the usual result
\begin{equation}
  \label{eq:instanton}
  \theta_k(t) = 4\arctan\exp\left[\sqrt{E_JE_C}\,(t-t')\right],
\end{equation}
where $t'$ is the arbitrary moment in time where the phase slip is
centered; we have assumed the limit of low temperature
$T\ll\sqrt{E_JE_C}$.  The action associated with this instanton trajectory
is $S_0=8\sqrt{E_J/E_C}$, \footnote{At non-zero self-capacitance of the
  grains one finds a correction $\tilde S \sim (E_J/\tilde E_C)^{1/2}\ln
  N$.  In the limit $N\to\infty$ it leads to the superconductor-insulator
  transition \cite{bradley}, but in finite systems our results for
  $I(\Phi)$ remain valid.  Furthermore, at $\tilde E_C\gg E_C$ the
  correction $\delta S$ is negligible unless $N$ is exponentially large.}.

The instantons account for the possibility of the system tunneling between
the different minima (\ref{eq:E_classical_chain}) of the potential energy
in action~(\ref{eq:action}).  The effect of the instantons on the ground
state energy can be accounted for by considering a tight-binding
Hamiltonian defined as
\begin{equation}
  \label{eq:tight-binding}
  H \psi_m= E_m \psi_m - Nv(\psi_{m-1}+\psi_{m+1}).
\end{equation}
Here $\psi_m$ is the amplitude of the system at the state $m$ with energy
$E_m$ given by Eq.~(\ref{eq:E_classical_chain}).  The hopping matrix
element is exponentially small, 
\begin{equation}
  \label{eq:hopping}
  v=\frac{4}{\sqrt\pi}(E_J^3E_C)^{1/4}
    \exp\left(-8\sqrt{\frac{E_J}{E_C}}\right), 
\end{equation}
where the exponent coincides with the instanton
action $S_0$, and the pre\-factor can be obtained by considering the
problem (\ref{eq:action}) in the case of a single junction (without the
constraint (\ref{eq:constraint})).  

It is important to note that for any $k$ the set of phases
$\theta_n=(\varphi-2\pi)/N + 2\pi\delta_{nk}$ describes the same physical
state of the chain.  Thus the hopping matrix elements due to the
instantons (\ref{eq:instanton}) in all the $N$ junctions must be summed
up, resulting in the additional factor of $N$ in the hopping term in the
Hamiltonian (\ref{eq:tight-binding}).

At $Nv\ll E_J/N$ the hopping term is small compared with the diagonal
matrix elements in the Hamiltonian (\ref{eq:tight-binding}).  Its effect
is most significant when $\Phi/\Phi_0$ is half-integer, and the energy
levels $E_m$ are degenerate, Fig.~\ref{fig:parabolas}(a).  In this regime
hopping gives rise to the level repulsion shown by the dashed line in
Fig.~\ref{fig:parabolas}(a) and the respective rounding of the sawtooth in
current, Fig.~\ref{fig:parabolas}(b).  The shape of the current steps is
given by
\begin{equation}
  \label{eq:current_step}
  I = \frac{2eE_J}{\hbar N}
      \left[\chi-\frac{\pi\chi}{\sqrt{\chi^2 +(vN^2/\pi E_J)^2}}\right],
\end{equation}
where $\chi=\varphi-\pi=2\pi(\Phi/\Phi_0-1/2)\ll1$.

In general hopping affects the spectrum of the Hamiltonian
(\ref{eq:tight-binding}) dramatically.  To find the ground state energy
$E(\varphi)$ it is more convenient to apply the Hamiltonian
(\ref{eq:tight-binding}) to the wavefunction in ``coordinate
representation,'' $\psi(x)=\sum_m \psi_m e^{i(2m-\varphi/\pi)x}$.  The
resulting Schr\"odinger equation then takes the form of Mathieu equation:
\begin{equation}
  \label{eq:mathieu}
  \psi''(x) + (a-2q\cos 2x)\psi(x)=0,
\end{equation}
where parameter $a=2NE/\pi^2E_J$ is proportional to the energy $E$, and
$q=N^2v/\pi^2E_J$.  The phase $\varphi$ enters the problem via a twisted
boundary condition $\psi(x+\pi)=e^{-i\varphi}\psi(x)$.  The regime of
strong hopping in the Hamiltonian (\ref{eq:tight-binding}) corresponds to
$q\gg1$.  In this case the dependence of the eigenvalue $a$ on the phase
$\varphi$ is exponentially small, $a=-16\sqrt{2/\pi}q^{3/4} e^{-4\sqrt
  q}\cos\varphi$, Ref.~\onlinecite{abramowitz}.  For the persistent
current $I=(2e/\hbar)dE/d\varphi$ we then find
\begin{equation}
  \label{eq:current_sinusoidal}
  I=16\sqrt{2N}\,\frac{e}{\hbar}(E_Jv^3)^{1/4}
     \exp\left(-\frac{4}{\pi}N\sqrt{\frac{v}{E_J}}\right)
    \sin\varphi.
\end{equation}

It is important to note that the Schr\"odinger equation for the
Hamiltonian (\ref{eq:tight-binding}) coincides with Eq.~(\ref{eq:mathieu})
at arbitrary hopping strength $q$.  Thus the problem of the flux
dependence of the persistent current reduces to the solving the well known
Mathieu equation (\ref{eq:mathieu}) with the appropriate twisted boundary
conditions. This conclusion is our main result.  The shape of the current
oscillations is controlled by a single parameter $q$.  As $q$ changes from
0 to $\infty$, the shape crosses over from sawtooth to the sinusoidal
behavior.

The Hamiltonian (\ref{eq:tight-binding}) has been rigorously derived for
the model (\ref{eq:action}) of a chain of coupled superconducting grains.
Although nanorings are not described by this model, the essential
low-energy physics of a superconducting state being destroyed by quantum
phase slips remains the same, and the Hamiltonian (\ref{eq:tight-binding})
and the respective Schr\"odinger equation (\ref{eq:mathieu}) are still
valid.

To apply the Hamiltonian (\ref{eq:tight-binding}) to superconducting
nanorings, one needs to relate the matrix elements $E_m$ and $Nv$ to the
parameters describing the nanowires, such as their length $L$, coherence
and localization lengths $\xi$ and $\lambda$, etc.  The diagonal matrix
elements $E_m$ are given by their classical values (\ref{eq:E_classical}).
The important difference between nanowires and chains of superconducting
grains appears in evaluating the hopping matrix element $Nv$.  In the
absence of tunneling barriers the quantum phase slips must involve
suppression of the superconducting gap $\Delta$ in the wire.  The typical
volume of space where the gap is suppressed is $V\sim A\xi$.  Given the
condensation energy per unit volume $\varepsilon\sim m^* k_F\Delta^2$ and
the typical correlation time $\tau\sim1/\Delta$, one can estimate the
phase slip action in the wire $S_0\sim \varepsilon V\tau
\sim\sqrt{\Delta/\delta_\lambda} \sim\lambda/\xi$.  A more rigorous
treatment \cite{zaikin,golubev} of the quantum phase-slips gives rise to
the same result, but unfortunately the numerical coefficient cannot be
obtained analytically.  Taking into account the obvious analogy between
the number of Josephson contacts $N$ and the length $L$ of the wire, we
conclude $Nv\propto Le^{-S_0}$, with $S_0\sim\lambda/\xi$.

Another feature of realistic nanowires not included in our model of a
chain of Josephson junctions is the mesoscopic fluctuations introduced by
the disorder.  For instance, since the instanton action depends on the
mean free path, the fluctuations of the random potential can lead to the
fluctuations of $S_0$.  This effect should lead to mesoscopic fluctuations
of the persistent current; the magnitude of the fluctuations is expected
to be small in relatively thick wires.

A more subtle effect of the random potential affects the relative phases
of instantons in different parts of the wire.  It can be explored within
our model by adding gates to the grains and applying random gate voltages.
It has been shown recently \cite{ioffe,averin} that the gate voltages give
rise to Aharonov-Casher interference effects for the instantons in
different junctions, and under certain conditions can suppress the effects
of the phase slips.

To account for the effect of the random gate potential on the persistent
current, we present the electrostatic charging energy of the chain as
\begin{equation}
  \label{eq:electrostatic}
  H_C=\sum_{nl}\frac12 [C^{-1}]_{nl}(Q_n-q_n)(Q_l-q_l),
\end{equation}
where $[C^{-1}]_{nl}$ is the matrix of inverse capacitances of the system,
$Q_n$ is the charge at the $n$-th grain.  Parameters $q_n$ are determined
by the gate voltages and have the meaning of the values of the charges
$Q_n$ at which the electrostatic energy is minimized.  

Using the fact that the grain charge $Q_n$ and phase $\phi_n$ are
conjugate variables and introducing $\theta_n = \phi_n - \phi_{n+1} +
\varphi/N$, one rederives the imaginary-time action (\ref{eq:action}) with
an additional term
\begin{equation}
  \label{eq:gate_action}
  \delta S= -i\sum_n p_n\dot\theta_n,
  \quad p_n=q_1+q_2+\dots+q_n,
\end{equation}
accounting for the gate voltages.  Since the instanton
(\ref{eq:instanton}) corresponds to a trajectory where the phase in
contact $k$ changes by $2\pi$, the respective contribution to the hopping
matrix element in the Hamiltonian (\ref{eq:tight-binding}) has an
additional phase $e^{i2\pi p_n}$.  Therefore for the total hopping matrix
element we obtain $v\sum_n e^{i2\pi p_n}$ instead of $Nv$, in agreement
with Ref.~\onlinecite{ioffe}.  If the random gate voltages are
sufficiently strong, so that $q_n\sim1$, the average of the hopping matrix
element in the Hamiltonian (\ref{eq:tight-binding}) vanishes: $\langle
v\sum_n e^{i2\pi p_n}\rangle=0$. Its typical values are of the order of
the mean square fluctuation, thus decreasing from $Nv$ in the regular case
to $\sim\sqrt Nv$ in the case of random charges.  In the limit of large
$N$, the same square-root dependence on the length of the chain will
result even from weak disorder. This suppression of the effect of phase
slips results in a higher value of the persistent current, which can be
estimated by replacing $v\to v/\sqrt N$ in Eqs.~(\ref{eq:current_step})
and (\ref{eq:current_sinusoidal}).  The most significant change in the
properties of the persistent current caused by the random gate potential
is that instead of $\ln I\propto-N$ we find a slower dependence $\ln
I\propto-N^{3/4}$ on the length of the chain at large $N$.

It is worth noting that the above result cannot be obtained from the
action averaged over the random gate potentials $q_n$.  Indeed, one can
easily see that $\langle \delta S\rangle=0$, thereby restoring our
previous results (\ref{eq:current_step}) and
(\ref{eq:current_sinusoidal}).  A term similar to (\ref{eq:gate_action})
also appears in the theory of the persistent current in disordered wires.
This term breaks the translational invariance and destroys the zero mode
of the instanton action.  Thus the conventional argument that the zero
mode gives rise to the prefactor of the instanton contributions to the
matrix elements proportional to the length of the system does not apply.
Consequently we expect the total matrix element in the Hamiltonian
(\ref{eq:tight-binding}) to behave as $\sqrt L \,e^{-S_0}$, leading to the
persistent current $\ln I \propto -L^{3/4}e^{-S_0/2}$.  Averaging of the
action over the disorder would restore the zero mode and lead to the
incorrect result $\ln I \propto -Le^{-S_0/2}$.

Our theoretical predictions can be tested in devices similar to those of
Refs.~\onlinecite{bezryadin1,bezryadin2}.  Although we discussed the
geometry of a nanoring of uniform thickness, all the conclusions are
applicable in the case of a straight nanowire with ends shorted by a bulk
superconductor.  In this geometry only the quantum phase slips in the
nanowire part should be considered, and thus the length of the system $L$
is that of the nanowire only.  The most interesting theoretical prediction
to be tested in such devices is the dependence of the persistent current
on the magnetic flux $\Phi$.  The dependence $I(\Phi)$ evolves from
sawtooth to the sinusoidal one, Fig.~\ref{fig:parabolas}(b), as the
nanowire becomes longer and/or thiner.  We expect that the experimentally
measured shape of the oscillations of $I(\Phi)$ can be fitted to the
function $da(\varphi)/d\varphi$, determined from the Mathieu equation
(\ref{eq:mathieu}).  The only free parameter in such a fit is the
effective phase slip strength $q$.

Another type of systems to which our predictions should apply is the
one-dimensional arrays of Josephson junctions.  Recently it became
possible to tune the Josephson coupling energy in such devices by external
magnetic field \cite{haviland}.  As a result, one should be able to study
the whole crossover from sawtooth to sinusoidal dependence of $I(\Phi)$ in
a single sample.

In conclusion, we have studied the persistent current in nanoscale
superconducting rings.  As the wire becomes thiner, the quantum phase
slips begin to suppress the current and change its flux dependence
$I(\Phi)$.  The latter can be found by solving Mathieu equation
(\ref{eq:mathieu}) with the appropriate boundary condition.  The
Aharonov-Casher interference suppresses the effects of the phase slips and
gives rise to the unusual length dependence $\ln I\propto -L^{3/4}$.

\begin{acknowledgments}
  KAM acknowledges the support of the Sloan Foundation and NSF
  Grant DMR-9974435. The work at the University of Minnesota is
  supported by NSF Grants DMR-9731756 and DMR-0120702.  The authors
  acknowledge the hospitality of Aspen Center for Physics and ITP at
  UC-Santa Barbara, where part of the work was performed.
\end{acknowledgments}

\end{document}